\documentclass[aps,prd,final,twocolumn,floats,floatfix,nofootinbib,10pt]{revtex4-1}

\usepackage{blindtext}  
\usepackage{graphicx}
\usepackage{amsmath}
\usepackage{amsfonts}
\usepackage{amssymb}
\usepackage{amstext}
\usepackage{mathtools}
\usepackage[english]{babel}
\usepackage{helvet}
\usepackage{slashed}
\usepackage{microtype}
\usepackage[pdftex]{hyperref}
\usepackage{cancel}

\begin{document}

\title{Non-relativistic limit for Higher Spin Fields and Planar Schroedinger Equation in 3D. }


\author{Abhijeet Dutta$^{\text{1,2}}$}

\address{1: BRAC University, Dhaka, Bangladesh}
\address{2: Tensor, Dhaka, Bangladesh. }

\date{\today}

\begin{abstract}
    Higher spin (HS) fields naturally occur in string theory, they are considered as a candidate for dark matter and may also appear as a collective excitation in condensed matter systems. In some cases one may study the HS fields in the non-relativistic settings. Thus, it is of interest to know the non-relativistic limit of HS fields and how to find the Schroedinger equation as the dynamical equation in this limit. In this paper, we consider the  non-relativistic limit of HS fields in Minkowskian spacetime in 3D. We work both at the level of equation of motion and action/Lagrangian density. We find the systematic procedures in both settings and show that they can be generalized to arbitrary HS fields.
\end{abstract}

\maketitle

\section{Introduction}
The study of higher spin fields and their non-relativistic limit in (2+1)D may be important for various fields of research such as collective excitations in the condensed matter systems, 3D massive gravity, non-relativistic holography and string theory \cite{bergshoeff-3}.

Our goal in this paper is to find a Planar Schroedinger Equation (SE) for higher spin  fields by finding the non-relativistic limit of a relativistic theory in the Minkowski spacetime in $(2+1)$D. Wigner little group for massive fields in (2+1)D=3D is $SO(2)$. The DoFs for the massive higher spin  fields can be counted from the Pauli-Fierz conditions. In 3D, the DoF for the massive higher spin  fields is two \cite{rakib}. So, we need to find systematic procedures to identify the correct DoFs of the higher spin  fields and then, use them to arrive at the planar SE in the non-relativistic limit.

Bergshoeff et al. have found the planar SE for spin-0 and spin-2 fields \cite{bergshoeff-2}. They have shown that for a real spin-0 field, we cannot find a dynamical equation by taking the non-relativistic limit $c \rightarrow  \infty$ of the Klein-Gordon(KG) equation. However, we can find a dynamical equation for a spin-0 field after taking the non-relativistic limit, if we consider a complex field instead of a real field. To be precise, for a scalar field $\varphi$ that satisfies KG equation, we define $\varphi(\Vec{x},t) = e^{\frac{-i}{\hbar}mc^{2}t} \Psi(\Vec{x},t)$, where $\Psi \in \mathbb{C}$. Therefore the complex field has a projective representation. The central extension of the Galilean algebra is proportional to mass, which is the reason why we have a projective representation \cite{steve}. Bergshoeff et al. have proposed a new null reduction ansatz, which they used to find the planar SE in 3D, starting from the massless Fronsdal equation in 4D. Bergshoeff et al. have also shown how to find the planar SE from the Lagrangian density of a real massive vector field in 3D using non-local field re-definitions\cite{bergshoeff-1}.

\vspace{1mm}

First we show how to find the planar SE for higher spin  bosonic and fermionic fields starting from their respective Fronsdal equations in 4D. We follow the procedure taken by Bergshoeff et al. \cite{bergshoeff-2} - we work with the light-cone gauge condition and use the null reduction to find the planar SE in 3D. We do it for spin- 1, 3, 1/2, 3/2 and 5/2 fields. By doing so, we demonstrate that the procedure can be carried out systematically for all the higher spin  bosonic and fermionic fields. Bergshoeff et al. in \cite{bergshoeff-1} mentions that with the use of null reduction, one can find planar SE for any integer spin in 3D. \\
Kuzenko et al. have worked out the transverse, gamma-traceless projectors for fermionic fields and transverse, traceless projectors for bosonic fields in 3D \cite{kuzenko}. We use the projectors to find the independent DoFs in the massive Pauli-Fierz Lagrangian density. We eliminate the auxiliary fields by using their respective equations of motion. Then we complexify the independent DoFs to a complex DoF and use the projective representation. After the substitution of the complex field in the Lagrangian density, we take the non-relativistic limit $ c \rightarrow \infty $ and find the non-relativistic Lagrangian density. By using the Euler-Lagrange equation for this non-relativistic Lagrangian density, we find the planar SE for the respective higher spin  bosonic field. We demonstrate the procedure by carrying out the calculations for spin-1, spin-2 and spin-3 bosonic fields. Similarly, one can find the planar SE for other higher spin  bosonic fields. For the fermionic higher spin  fields, we use the projectors to find the positive and negative helicity fields. Both the projected fields satisfy the Dirac equation. Therefore, they also satisfy the Klein-Gordon(KG) equation. We take a linear combination of the independent DoFs. Then similar to the bosonic fields, we use the projective representation of the linear combination in the KG equation. Then taking the non-relativistic limit $c \rightarrow \infty$, we arrive at the planar SE. We show the procedure for spin-1/2 and spin-3/2 fields. The procedure can be carried out systematically for all the higher spin  fermionic fields. \\
We also give another Lagrangian density approach that is based on Kaluza-Klein reducing a 3+1 dimensional Lagrangian to a 3 dimensional Lagrangian. We then gauge fix the 3D Lagrangian and find a diagonal kinetic term for the independent dofs. Henceforth, we follow the procedure given in Bergshoeff et al. \cite{bergshoeff-2} for spin-$0$ and take the $c \longrightarrow \infty$ limit to find the planar Schr\"odinger equation.

\section{ Equation of motion Approach }

Here we follow the approach taken by Bergshoeff et al.\cite{bergshoeff-2} for the spin-2 field. We show that the procedure generalizes for arbitrary higher spin fields.\\
We start with a massless theory in 4D and use the null reduction ansatz proposed by Bergshoeff et al \cite{bergshoeff-2} to find the planar SE in 3D. First we consider spin-1 field and then we consider spin-3 field. Bergshoeff et al. have derived the planar SE for spin-2 field \cite{bergshoeff-2}.\\ 

We work with light-cone coordinates $x^{m} = (x^{+},x^{-},x^{I})$ where $(I=1,2)$, $x^{\pm} = \frac{x^{3} \pm x^{0}}{\sqrt{2}} $ and set $c=1$. \\

\textbf{Spin-1:} We start with the Fronsdal equation for a massless spin-1 field $\varphi_{m}$ in 4D:- 
\begin{align}
    \Box \varphi_{m} - \partial_{m}(\partial \cdot \varphi) = 0 \label{spin-1}
\end{align}

We impose the light-cone gauge condition:-
\begin{align}
    \varphi_{-} = 0 \label{phi minus}
\end{align}

We write eq.\eqref{spin-1} for $\varphi_{-}$ and use the light-cone gauge condition eq.\eqref{phi minus} to find:-
\begin{equation}
    \partial \cdot \varphi = 0 \label{spin-1 transverse}
\end{equation}

This is the transversality condition for spin-1. \\

The transversality condition eq.\eqref{spin-1 transverse} together with the light-cone gauge condition eq.\eqref{phi minus} implies the following subsidiary equation for the auxiliary variable $\varphi_{+}$:-
\begin{align}
    \partial_{+} \varphi_{+} = - \partial_{I} \varphi_{I} \label{phi+}
\end{align}

Now, we define a complex field $\Psi[1]$ by combining the two real DoFs of spin-1:-
\begin{align}
    \Psi [1] = \varphi_{1} + i \ \varphi_{2}
\end{align}

We also complexify the spatial coordinates, $z = x^{1} + i \ x^{2} $ and denote the complex coordinate as $z$. Working with complexified variables enables us to express the subsidiary equation(s) in a compact form and to employ the new null reduction of Bergshoeff et al. \\

We can write the subsidiary condition eq.\eqref{phi+} using the complex field $\Psi[1]$ as:- 
\begin{align}
    \partial_{+} \varphi_{+} = - \mathbb{R}(\partial \Psi[1])
\end{align}

where $\frac{\partial}{\partial z} := \partial $. \\

Following Bergshoeff et al. \cite{bergshoeff-2}, we define the null reduction in the following way:-
\begin{align}
    \partial_{-} \Psi[a] = \bigg(\frac{i m}{\hbar} \bigg) \Psi[a],  \  \  a = 1 \label{null reduction-1}
\end{align}

From eq.\eqref{spin-1} with the transversality condition eq.\eqref{spin-1 transverse}, we write the wave equation satisfied by the complex field as:-
\begin{align} 
     2 \partial_{+} \partial_{-} \Psi [1] = - (\partial^{2}_{1} + \partial^2_{2}) \Psi [1]
\end{align}

Then using the null reduction eq.\eqref{null reduction-1} we find the planar SE for the complex field $\Psi[1]$:- 
\begin{align}
    i  \hbar  \Dot{\Psi}[1] = - \bigg(\frac{\hbar^{2}}{2m} \bigg) \nabla^{2} \Psi [1]
\end{align}


where $\Dot{\Psi}[1] := \partial_{+} \Psi [1] $. \\

\textbf{Spin-3:} We consider a totally symmetric rank-3 tensor $\varphi_{mnp} $ for the spin-3 field where $m,n,p = 0,1,2,3$. \\

The Fronsdal equation for a massless  spin-3 field in 4D is given by :- 
\begin{equation}
   \begin{aligned}
    \Box \varphi_{mnp} & - (\partial_{m} \partial \cdot \varphi_{np} + \partial_{p} \partial \cdot \varphi_{mn} + \partial_{n} \partial \cdot \varphi_{pm} ) \\ 
    & + (\partial_{m} \partial_{n} \varphi'_{p} + \partial_{p} \partial_{m} \varphi'_{n}  + \partial_{n} \partial_{p} \varphi'_{m}) = 0 \\
    \label{fronsdal}
\end{aligned}
\end{equation}

\vspace*{-1.25\baselineskip}

where $\varphi'_{p}$ is the trace of $\varphi_{mnp}$, defined by $\varphi'_{p} = \eta^{mn} \varphi_{mnp} $. \\

We impose the light-cone gauge condition:-

\vspace*{-1.25\baselineskip}

\begin{align}
    \varphi_{-np} = 0
    \label{gauge}
\end{align}

Using the light-cone Minkowski metric we find the trace of $\varphi_{mnp}$ to be:-

\vspace*{-1.25\baselineskip}

\begin{align}
    \varphi'_{p} = \eta^{mn} \varphi_{mnp} = \varphi_{IIp}
    \label{trace1}
\end{align}

We define, $\varphi_{np} = \eta^{rq} \partial_{r} \varphi_{qnp} $. Expanding the contractions we find:-

\vspace*{-1.25\baselineskip}

\begin{align}
    \varphi_{np} = \partial_{-} \varphi_{+np} + \partial_{K} \varphi_{Knp}
    \label{trace2}
\end{align}
*Captalized letters correspond to spatial coordinates $1,2$. \\ 

By plugging in the values of n and p in eq.\eqref{trace2}, we find the following set of equations:-  

\vspace*{-1.25\baselineskip}
 
 \begin{align}
     \varphi_{++} & = \partial_{-} \varphi_{+++} + \partial_{K} \varphi_{K++}  \label{++} \\ 
     \varphi_{+I} & = \partial_{-} \varphi_{++I} + \partial_{K} \varphi_{K+I} \label{+I} \\
     \varphi_{IJ} & = \partial_{-} \varphi_{+IJ} + \partial_{K} \varphi_{KIJ}  \label{IJ}
 \end{align}

From the Fronsdal equation \eqref{fronsdal} for $\varphi_{mn-}$ and the light-cone gauge condition \eqref{gauge} we get:- 

\vspace*{-1.25\baselineskip}

\begin{align}
    \partial_{-} (\varphi_{mn} + \partial_{m}\varphi'_{n} + + \partial_{n}\varphi'_{m}) = 0
    \label{TT}
\end{align}

The equation \eqref{TT} implies:-

\vspace*{-1.25\baselineskip}

\begin{align}
    \varphi_{mn} &= 0 \label{transverse}\\
    \varphi'_{m} &= \varphi_{IIp} = 0 \label{traceless}
\end{align}

where the eq.\eqref{transverse} can be referred to as the transversality condition and the eq.\eqref{traceless} as the tracelessness condition. \\ 

By using the transversality condition eq.\eqref{transverse} in the equations eq.\eqref{++}, eq.\eqref{+I} and eq.\eqref{IJ}, we find:-

\vspace*{-1.25\baselineskip}
 
 \begin{align}
      \partial_{-} \varphi_{+++} & = - \partial_{K} \varphi_{K++} \label{subsidiary1} \\ 
     \partial_{-} \varphi_{++I} & =  - \partial_{K} \varphi_{K+I}  \label{subsidiary2} \\
     \partial_{-} \varphi_{+IJ} & =  - \partial_{K} \varphi_{KIJ} \label{subsidiary3}
 \end{align}

Now, we define two complex fields in the following way:-

\vspace*{-1.25\baselineskip}

\begin{align}
    \Psi[1] &= \varphi_{1++} + i \ \varphi_{2++} \label{psi1} \\
    \Psi[2] &= \varphi_{11+} + i \ \varphi_{12+} \label{psi2}
\end{align}


Similar to our approach in spin-1, We also complexify the spatial coordinates, $z = x^{1} + i \ x^{2} $. \\

\noindent Now, eq.\eqref{subsidiary1}, eq.\eqref{subsidiary2} and eq.\eqref{subsidiary3} respectively become:-

\vspace*{-1.25\baselineskip}

\begin{align}
    \partial_{-} \varphi_{+++} &= - \mathbb{R} (\partial \Psi[1] ) \label{subsidiary1*}  \\
    \partial_{-} \Psi[1] &= - \partial \Psi[2]  \label{subsidiary2*} \\
    \partial_{-} \Psi[2] &= - ( \partial \varphi_{111} + i \ \partial \varphi_{112} )  \label{subsidiary3*}
\end{align}

where $\partial$ is the derivative with respect to the complex z-coordinate. \\ 

From eq.\eqref{subsidiary1*}, eq.\eqref{subsidiary2*} and eq.\eqref{subsidiary3*}, we find $ \Psi[2] $ as the independent complex DoF which contains the two real DoFs $ \varphi_{11+} $ and $ \varphi_{12+} $. \\

In accordance with Bergshoeff et al.\cite{bergshoeff-2}, we define the null reduction:-
\begin{align}
    \partial_{-} \Psi[a] = \bigg(\frac{i m}{\hbar} \bigg) \Psi[a],  \  \  a = 1,2 \label{null reduction}
\end{align}

By using the null reduction eq.\eqref{null reduction}, we can write eq.\eqref{subsidiary2*} as:- 
 \begin{align}
     \Psi [1] = \bigg(\frac{i \hbar}{m} \bigg) \ \partial \Psi[2] 
 \end{align}

From the Fronsdal equation eq. \eqref{fronsdal}, with the transversality eq\eqref{transverse} and tracelessness eq.\eqref{traceless} conditions, We write the wave equation for the complex DoF $\Psi [2]$ as:- 
\begin{align}
    2 \ \partial_{+} \partial_{-} \Psi [2] &= - (\partial^{2}_{1} + \partial^{2}_{2} ) \  \Psi [2]
\end{align}
Thereupon, using the null reduction eq.\eqref{null reduction}, we find that $\Psi [2]$ satisfies the planar SE:- 
\begin{align}
    i  \hbar  \Dot{\Psi}[2] = - \bigg(\frac{\hbar^{2}}{2m} \bigg) \nabla^{2} \Psi [2]
\end{align}

\vspace*{-1\baselineskip}

where $\Dot{\Psi}[2] := \partial_{+} \Psi [2] $. \\

The procedure can be systematically carried out for higher spin $s>3$ bosonic fields. \\

Now, we show how we can carry out the null-reduction procedure for the fermionic higher spin  fields. \\

\textbf{Spin-1/2:} The procedure is straightforward for the spin-1/2 field. As it satisfies the massless Dirac equation in 4D, therefore it also satisfies the wave equation. Massive Dirac spinor in 3D has one complex DoF or two real DoFs. We use the null reduction\cite{bergshoeff-2} ansatz to find the planar SE for the spin-1/2 field. \\

\textbf{Spin-3/2:}
The Fronsdal equation for a massless spin-3/2 field, $\Psi_{m}$ in 4D is:-
\begin{align}
    \cancel{\partial} \Psi_{m} - \partial_{m} \cancel{\Psi}  = 0 \label{spin-3/2}
\end{align}

We impose the light-cone gauge condition:-

\vspace*{-1.25\baselineskip}

\begin{align}
    \Psi_{-} = 0 \label{lcg-3/2}
\end{align}

We implement the light-cone gauge condition eq.\eqref{lcg-3/2} in the Fronsdal equation eq.\eqref{spin-3/2} for $\Psi_{-}$. Whereupon, we find the gamma-tracelessness condition:-
\begin{align}
      \cancel{\Psi}  = 0 \label{gamma-traceless-3/2}
\end{align}

Now we take the divergence of equation eq.\eqref{spin-3/2} and use eq.\eqref{gamma-traceless-3/2} to find the divergenceless condition:-
\begin{align}
    \partial \cdot \Psi = 0 \label{divergenceless-3/2}
\end{align}

By using the subsidiary equations eq.\eqref{lcg-3/2}, eq.\eqref{gamma-traceless-3/2} and eq.\eqref{divergenceless-3/2}, we can eliminate the auxiliary variables. We choose $\Psi_{1}$ to be the complex independent DoF for the spin-3/2 field. \\

Using the gamma-tracelessness condition eq.\eqref{lcg-3/2}, we find $\Psi_{1}$ satisfies the massless Dirac equation:-
\begin{align}
    \cancel{\partial} \Psi_{1} = 0
\end{align}

Therefore, $\Psi_{1}$ also satisfies the wave equation:-
\begin{align}
     2 \ \partial_{+} \partial_{-} \Psi_{1} &= - (\partial^{2}_{1} + \partial^{2}_{2} ) \  \Psi_{1} \label{wave-3/2}
\end{align}
\
Now we write the null reduction ansatz for this component of the spin-3/2 field :-

\vspace*{-1.25\baselineskip}

\begin{align}
    \partial_{-} \Psi_{1} = \bigg(\frac{i m}{\hbar} \bigg) \Psi_{1} \label{null reduction-3/2}
\end{align}

By means of the null reduction eq.\eqref{null reduction-3/2}, from eq.\eqref{wave-3/2} we find the planar SE for the spin-3/2 field:-
\begin{align}
    i  \hbar  \Dot{\Psi}_{1} = - \bigg(\frac{\hbar^{2}}{2m} \bigg) \nabla^{2} \Psi_{1}
\end{align}

Now we sketch the procedure for spin-5/2 field. \\

\textbf{Spin-5/2:} The Fronsdal equation for the massless spin-5/2 field, $\Psi_{mn}$ in 4D is given by:-
\begin{align}
    \cancel{\partial} \Psi_{mn} - \partial_{m}\cancel{\Psi}_{n} - \partial_{n}\cancel{\Psi}_{m} = 0 \label{spin-5/2}
\end{align}

The light-cone gauge condition is:-
\begin{align}
    \Psi_{-n} =0 \label{lcg-5/2}
\end{align}

We write the Fronsdal equation eq.\eqref{spin-5/2} for the component $\Psi_{-n}$ with the light-cone gauge condition eq.\eqref{lcg-5/2}. We find the gamma-traceless condition :-
\begin{align}
    \cancel{\Psi_{n}} = 0 \label{gamma-5/2}
\end{align}

By taking the divergence of the Fronsdal equation eq.\eqref{spin-5/2} with the gamma-traceless condition eq.\eqref{gamma-5/2}, we find the divergenceless condition:-
\begin{align}
    \partial \cdot \Psi_{n} = 0 \label{divergenceless-5/2}
\end{align}

Now, We can eliminate the auxiliary variables by the help of the equations eq.\eqref{lcg-5/2}. eq.\eqref{gamma-5/2} and eq.\eqref{divergenceless-5/2}, and find the one independent complex DoF. We choose this DoF to be $\Psi_{11}$ and write the null reduction\cite{bergshoeff-2} ansatz for it. Then making use of the null reduction ansatz we can find the planar SE for the spin-5/2 field. \\

Henceforward the procedure can be carried out systematically for all the higher spin  fermionic fields. This concludes the equation of motion approach. Now we look at the Lagrangian density approach. 

\section{Lagrangian density TT Projecctor Approach}

We can also find the planar SE in 3D by working from the Lagrangian density of massive higher spin fields in 3D Minkowski spacetime. We use the transverse, traceless projectors for the bosonic higher spin fields and transverse, gamma-traceless projectors for the fermionic higher spin fields. They are worked out by Kuzenko et al \cite{kuzenko}. We use the cartesian coordinate system $x^{a} = (x^{0}, x^{1}, x^{2})$. Minkowski metric in this system is $\eta_{ab} = diag(-c^{2},1,1)$ and $\sqrt{-det\eta}=c$, where $c$ is the speed of light.

First we consider the bosonic higher spin fields.\\

\textbf{Bosonic Higher Spin Fields:}
We show the procedure explicitly for spin-1 and spin-2 fields. From which it will become clear that the procedure can be systematically carried out for arbitrary higher spin bosonic fields. \\

\textbf{Spin-1:} The transverse projector \cite{kuzenko} for spin-1 field is given by :-
\begin{align}
    \Pi^{[1]}{^b_a} = \frac{1}{\Box} (\Box \eta^{b}_{a} - \partial_{a} \partial^{b})
\end{align}

Let us define the transverse spin-1 field $h^{T}_{a}$ by \cite{kuzenko}:-
\begin{align}
    h^{T}_{a} &= \Pi^{[1]}{^b_a} \ h_{b} \\
=>   h^{T}_{a}   &=  \bigg( h_{a} - \frac{\partial_{a} (\partial \cdot h)}{\Box}\bigg) \label{hT-1}
\end{align}

We can easily verify that our transverse spin-1 field satisfies the transversality condition: $\partial \cdot h = 0$. This condition reduces the DoF by one. \\

Now we write down the Lagrangian density for the transverse spin-1 field $h^{T}_{a}$ in 3D Minkowski spacetime:- 

\vspace*{-1.25\baselineskip}

\begin{align}
    c^{-1} \mathcal{L} = - \sqrt{-det\eta} \ \frac{1}{4} F^{T}_{ab} F^{Tab} - \sqrt{-det\eta} \  \frac{(mc)^2}{2} h^{Ta} h^{T}_{a}  \
\end{align}
where $F^{T}_{ab} = \partial_{a} h^{T}_{b} - \partial_{b} h^{T}_{a}$. \\

By performing integration by parts and using the transversality condition we find:-
\begin{equation}
    \begin{split}
        & c^{-1} \mathcal{L} = \frac{1}{2} \bigg( \frac{-1}{c^{2}} h^{T}_{0} \Box h^{T}_{0} + h^{T}_{i} \Box h^{T}_{i} + \frac{(mc)^2}{c^2}  h^{T}_{0} h^{T}_{0} \\
        & - (mc)^2 h^{T}_{i} h^{T}_{i}   \bigg) ; \hspace{1.5mm}  \text{where i=1,2} \label{Lagrangian-1}
    \end{split}
\end{equation}

Because of the transversality constraint, we know that one of the components of the transverse spin-1 field is auxiliary. We choose the auxiliary variable such that we get a finite non-relativistic Lagrangian density as we take the non-relativistic limit $c \rightarrow \infty$. By inspection, we choose $h^{T}_{0}$ as the auxiliary component. We can eliminate $h^{T}_{0}$ by using its equation of motion:-
\begin{align}
    (\Box - (mc)^2) h^{T}_{0} = 0
\end{align}

After eliminating $h^{T}_{0}$, we are left with the following Lagrangian density:-
\begin{align}
    c^{-1} \mathcal{L} = \frac{-1}{2c^{2}} h^{T}_{i} \partial^{2}_{t} h^{T}_{i} + \frac{1}{2} h^{T}_{i} \nabla^{2} h^{T}_{i} - \frac{1}{2} (mc)^{2} h^{T}_{i} h^{T}_{i}
\end{align}

Now, we define a complex field,
\begin{align}
    H = ( h^{T}_{1} + i \ h^{T}_{2} )/ \sqrt{2}
    \label{s2-100}
\end{align}

We write the Lagrangian density eq.\eqref{Lagrangian-1} with the complex field $H$:-
\begin{align}
    c^{-1} \mathcal{L} = \frac{1}{c^{2}} |\Dot{H}|^{2} +  \Bar{H} \nabla^{2} H -  (mc)^{2} |H|^{2} \label{H-1}
\end{align}

Following Bergshoeff et al. \cite{bergshoeff-1}, we use the projective representation of $H$ :-
\begin{align}
    H(x,t) = e^{-imc^2t} \Psi(x.t) \label{projective-1}
\end{align}
\indent where $\Psi(x,t)$ is a complex function and we set $\hbar = 1$. \\

Plugging in the expression of $H$, eq.\eqref{projective-1} in eq.\eqref{H-1} we get:-
\begin{align}
     c^{-1} \mathcal{L} = 2 i m \Bar{\Psi} \Dot{\Psi} + \Bar{\Psi} \nabla^{2} \Psi
\end{align}

Now we take the non-relativistic limit $c \rightarrow \infty$ and find the non-relativistic Lagrangian density:-
\begin{align}
     c^{-1} \mathcal{L}_{NR} = 2 i m \Bar{\Psi} \Dot{\Psi} + \Bar{\Psi} \nabla^{2} \Psi
\end{align}

The Euler-Lagrange equation for this Lagrangian density gives the planar SE for the spin-1 field:-
\begin{align}
    i  \Dot{\Psi} = \frac{-1}{2 m} \nabla^{2} \Psi
\end{align}

Now we work out the planar SE for the spin-2 field. \\

\textbf{Spin-2:} We represent the spin-2 field as a totally symmetric, traceless rank-2 tensor $h_{ab}$. \\

With the help of the transverse, traceless projector \cite{kuzenko} for the spin-2 field, we write the transverse, traceless spin-2 field $h^{T}_{ab}$ as :-
\begin{equation}
    \begin{split}
        h^{T}_{ab} &= \Pi^{[2]cd}_{ab} \ h_{cd}\\
        & = \bigg( h_{ab} - \frac{2}{\Box} \partial^{c} \partial_{(a}h_{b)c} + \frac{1}{2\Box} \eta_{ab} \partial^{c} \partial^{d} h_{cd} \\
        & + \frac{1}{2\Box^{2}} \partial_{a} \partial_{b} \partial^{c} \partial^{d} h_{cd} \bigg)
    \end{split}
\end{equation}

The projected field satisfies the namesake transversality and tracelessness conditions:-
\begin{align}
    \partial^{a} h^{T}_{ab} &= 0 \label{transverse-2} \\
    h^{Ta}_{a} &= 0 \label{traceless-2}
\end{align}

Now, we write the Pauli-Fierz Lagrangian density for the transverse, traceless spin-2 field \cite{rakib}:-
\begin{equation}
    \begin{split}
         c^{-1} \mathcal{L} &= \sqrt{-det\eta} \ \bigg( - \frac{1}{2} (\partial_{a}  h^{T}_{bc})^{2} +  (\partial \cdot  h^{T}_{b})^{2} + \frac{1}{2} (\partial_{a}  h^{Tb}_{b})^{2} \\
        & - (\partial \cdot  h^{T}_{a}) (\partial^{a} h^{Tb}_{b}) - \frac{1}{2} (mc)^{2} [h^{T2}_{ab} - h^{Tb2}_{b} ] \bigg)  \label{fp-2}
    \end{split}
\end{equation}

As  $h^{T}_{ab}$ satisfies the transversality eq.\eqref{transverse-2} and traceless eq.\eqref{traceless-2} conditions, its Pauli-Fierz equation becomes:-
\begin{align}
    c^{-1} \ \mathcal{L} &= -\frac{1}{2} (\partial_{a}  h^{T}_{bc})^{2} - \frac{1}{2} (mc)^{2} h^{T2}_{ab} 
\end{align}

A symmetric rank-2 field in 3D has six DoF. However, we have four constraints coming from the transversality eq.\eqref{transverse-2} and tracelessness eq.\eqref{traceless-2} conditions. Therefore, we have two independent components and we can solve the other auxiliary components in terms of them. Again we have to choose the independent components such that they remain finite as we take the non-relativistic limit $c \rightarrow \infty$. With these remarks, first we eliminate $h^{T}_{00}$ and $h^{T}_{0i}$ components by using their equation of motion. Hence, we're left with:-
\begin{align}
    c^{-1} \ \mathcal{L} &= \frac{1}{2c^{2}} \Dot{h}^{T2}_{ij} - \frac{1}{2} (\partial_{k} h^{T}_{ij})^{2} - \frac{(mc)^{2}}{2} h^{T2}_{ij} 
\end{align}

We have eliminated three auxiliary variables. We have to eliminate one more. $h^{T}_{ij}$ has three components: $h^{T}_{11}$, $h^{T}_{12}$ and $h^{T}_{22}$. We eliminate $h^{T}_{22}$ by using its equation of motion. We now have the Lagrangian density containing terms of only the two independent components:-
\begin{equation}
\begin{split}
    c^{-1} \ \mathcal{L} &= \frac{1}{2c^{2}} \Dot{h}^{T2}_{11} + \frac{1}{2c^{2}} \Dot{h}^{T2}_{12} + \frac{1}{2c^{2}} h^{T}_{11} \nabla^{2} h^{T}_{11} \\
        & +  \frac{1}{2c^{2}} h^{T}_{12} \nabla^{2} h^{T}_{12} - \frac{(mc)^{2}}{2} h^{T2}_{11} - \frac{(mc)^{2}}{2} h^{T2}_{12} 
        \label{Lagrangian-2} 
        \end{split}
\end{equation}
Now, we define a complex field $H$ by taking a complex combination of the two real DoFs $h^{T}_{11}$ and $h^{T}_{12}$:-
\begin{align}
    H = (h^{T}_{11} + i \ h^{T}_{12})/\sqrt{2}
\end{align}
We write the Lagrangian density eq.\eqref{Lagrangian-2} in terms of $H$:-
\begin{align}
    c^{-1} \mathcal{L} = \frac{1}{c^{2}} |\Dot{H}|^{2} +  \Bar{H} \nabla^{2} H -  (mc)^{2} |H|^{2} \label{H-2}
\end{align}
Then we express $H(\Vec{x},t)$ in terms of another complex variable $\Psi(\Vec{x},t)$ \cite{bergshoeff-1}:-
\begin{align}
    H(\Vec{x},t) = e^{-imc^{2}t} \Psi(\Vec{x},t) 
\end{align}
By plugging in this expression for $H(\Vec{x},t)$ in the Lagrangian density eq.\eqref{H-2}, we take the non-relativistic limit $c \rightarrow \infty$ and find the planar SE in 3D for the spin-2 field:-
\begin{align}
    i  \Dot{\Psi} = \frac{-1}{2 m} \nabla^{2} \Psi
\end{align}
From the calculations of spin-1 and spin-2 field, we find that the procedure is systematic and can be carried out for all the higher spin bosonic fields. 

Now, we show the procedure for the fermionic higher spin fields. \\


\textbf{Fermionic Higher Spin Fields:} Kuzenko et al. have shown that higher spin fields with spin $= \frac{n}{2} ; n \in \mathbb{N}$ can be projected into positive$+\frac{n}{2}$ and negative $-\frac{n}{2}$ helicity fields in 3D \cite{kuzenko}. They have also explicitly constructed the transverse, gamma-traceless projectors for spin-3/2 and spin-5/2 fields \cite{kuzenko}. \\

Here, we start with the spin-1/2 field and then with the help of projector, we show how it works for the spin-3/2 field. The procedure can be carried out for all the higher spin fields accordingly. \\

\textbf{Spin-1/2:} We represent the spin-1/2 field by the Dirac spinor, $\psi$. Following Kuzenko et al. \cite{kuzenko}, we can project the Dirac spinor in the positive and negative helicity fields via the respective helicity projectors. They are related to each other via Dirac conjugation.
\begin{align}
    \Pi^{[+]} \psi = \psi^{[+]} \ ;\
    \Pi^{[-]} \psi = \psi^{[-]} \label{projector-1/2}
\end{align}

The Lagrangian density for both the positive and negative helicity fields is given by:-
\begin{align}
    c^{-1} \mathcal{L} = \Bar{\psi}^{[+]} ( i \cancel{\partial} - mc ) \psi^{[+]} + \Bar{\psi}^{[-]} ( i \cancel{\partial} - mc ) \psi^{[-]}
\end{align}

From the Euler-Lagrange equations, they eq.\eqref{projector-1/2} both satisfy the Dirac equation:-
\begin{align}
    ( i \cancel{\partial} - mc ) \psi^{[+]} &= 0 \\
    ( i \cancel{\partial} - mc ) \psi^{[-]} &= 0
\end{align}
Therefore, they also satisfy the KG equation. We define a complex field, $\psi = ( \psi^{[+]} +  \psi^{[-]} )$ by combining $\Psi^{[+]}$ and $\Psi^{[-]}$. And write the KG equation for the combined complex field $\psi$:-
\begin{align}
    \frac{-1}{c^{2}} \ddot{\psi} + \nabla^{2} \psi -  (mc)^{2} \psi = 0 \label{eom-1/2}
\end{align}

Then we write the projective representation $\psi(x,t) = e^{-imc^2t} \Psi(x,t) $ as we did for the bosonic fields and plug it into the KG equation eq.\eqref{eom-1/2}. Henceforth, we take the non-relativistic limit $c \rightarrow \infty$ and find the planar SE for the spin-1/2 field. \\

For the spin-1/2 field, we could have just worked with the Dirac spinor instead of projecting it into helicity fields. However we need to identify the DoFs correctly and follow a procedure that can be generalized to arbitrary higher spin fermionic fields.\\

Now we find the planar SE for spin-3/2 field. \\

\textbf{Spin-3/2:} We represent the spin-3/2 field by $\psi_{a}$ (where $a$ is the vector index), which is gamma-traceless $ \gamma \cdot \psi = 0 $. We suppress the spinor index. We can project the spin-3/2 field into positive and negative helicity fields \cite{kuzenko}. We can also combine the projectors and construct a single projector $ \Pi^{[3/2]} = \Pi^{[3/2][+]} + \Pi^{[3/2][-]} $. By acting the projector on the spin-3/2 field,  we find \cite{kuzenko}:-
\begin{align}
    \Pi^{[3/2]} \psi_{a} = \frac{1}{\Box}\bigg( \psi_{a} - \partial_{a} \partial^{b} \psi_{b} - \frac{1}{2} \epsilon_{abc} \gamma^{b} \partial^{c} \partial^{d} \psi_{d} \bigg) \label{projector-3/2}
\end{align}

We define the projected spin-3/2 field as $  \psi^{T}_{a} := \Pi^{[3/2]}  \psi_{a} $. One can verify that $\psi^{T}_{a}$ in eq.\eqref{projector-3/2} satisfies the transversality and gamma-tracelessness conditions. \\

\vspace*{-0.5\baselineskip}

The Lagrangian density for the transverse, gamma-traceless spin-3/2 field $ \psi^{T}_{a}$ is given by:-
\begin{align}
    c^{-1} \mathcal{L} = - \Bar{\psi}^{T}_{a} \bigg(\gamma^{abc} \partial_{b} - imc \gamma^{ac} \bigg) \psi^{T}_{c}
\end{align}

By using $\gamma^{abc} = \gamma^{a} \gamma^{bc} - 2 i \eta^{a[b} \gamma^{c]}$ and $\gamma^{ac} = i ( \gamma^{a} \gamma^{c} -  \eta^{ac} )  $ (where $ X^{[ab]} = \frac{X^{a}X^{b}-X^{b}X^{a}}{2} $) we find:-
\begin{align}
    c^{-1} \mathcal{L} = - \Bar{\psi}^{T}_{a} ( i \cancel{\partial} - mc ) \psi^{Ta}
\end{align}
- which is the Dirac Lagrangian density for the transverse, gamma-traceless spin-3/2 field. Now we find the equation of motion which is the Dirac equation for $\psi^{T}_{a}$ from the Euler-Lagrange equation. From which we can find the KG equation for  $\psi^{T}_{a}$ and follow the similar procedure as we did for the spin-1/2 field.

Thus, the procedure is clear for all the higher spin fermionic fields. We use the projector to find the transverse, gamma-traceless field. Then we write the Lagrangian density for the respective field. With some gamma matrix manipulations and use of the transversality and gamma-tracelessness conditions give us the Dirac Lagrangian density for the respective field. And the rest of the calculation is identical to what we did for the spin-1/2 field. 

\section{Lagrangian Density KK reduction approach}
In this section, we describe another systematic method, the Kaluza-Klein (KK) reduction, for finding the planar SE using the Lagrangian density, which can be generalized to any higher-spin bosonic or fermionic field. We work it out for spin-$2$, spin-$3$ and spin-$3/2$. \\

\textbf{Spin-2:} The Fronsdal Lagrangian for a massless real totally symmetric spin$=2$ field $\Phi_{AB}$ in $D+1$ dimensions reads \cite{rakib} \cite{porrati} \cite{porrati-2}:
\begin{align}
        & {c^{-1} \mathcal{L}}=-\frac{1}{2}\left(\partial_E \Phi_{A B}\right)^2+ \left(\partial_A \Phi^{A B}\right)^2+  \ \Phi^{\prime} \partial_A \partial_B \ \Phi^{A B} +\frac{1}{2}\left(\partial_A \ \Phi^{\prime}\right)^2
        \label{2.1}
\end{align}
where $\Phi^{\prime} := \eta^{AB} \Phi_{AB}$ is the trace of the real totally symmetric spin$=2$ field $\Phi_{AB}$ . \\

The gauge symmetry of this Lagrangian is:-
\begin{align}
    \delta \Phi_{AB} = 2! \ \partial_{(A}\Lambda_{B)}
\end{align}
where, $\Lambda_{A}$ is the gauge parameter.\\
The gauge fixing term for the D+1 dimensional Lagrangian reads:
\begin{align}
    c^{-1} \mathcal{L} &= - \left( \partial^{A}  \Phi_{A B} - \frac{1}{2} \partial_{B} \Phi^{\prime} \right)^{2}
\end{align}

Incorporating the gauge fixing term in the Lagrangian of eq.(\ref{2.1}), we find:-
\begin{align}
    c^{-1} \mathcal{L} &= \frac{1}{2} \Phi_{A B} \square \Phi^{A B} - \frac{1}{4}  \Phi^{\prime} \square  \Phi^{\prime}
    \label{2.5}
\end{align}
In the above equation, $\square$ is a $D+1$ dimensional d'Alembertian. \\

The KK dimensional reduction ansatz is:
\begin{align}
    \Phi_{A B}\left(x^\mu, y\right)=  \sqrt{\frac{m}{2 \pi}} \frac{1}{\sqrt{2}}  \left[\varphi_{A B}\left(x^\mu\right) e^{i m y}+c \cdot c \cdot\right]
\end{align}
where we compactify the $y :=x_{D+1}$ spatial dimension on a circle of radius = $1/m$. \\

We define the fields in D dimensions as: $\varphi_{\mu \nu} =: h_{{\mu \nu}} , \ -i\varphi_{\mu y} =: B_{\mu}, \
    -\varphi_{y y} =: \phi $.
    Here $h_{\mu \nu}$ is a spin-$2$ field, $B_{\mu}$ is a spin-$1$ field and $\phi$ is a spin-$0$ field in dimension D. \\
We also write the KK ansatz for the gauge parameter $\Lambda_{A}$ as:- 
\begin{align}
    \Lambda_{A}\left(x^\mu, y\right)=  \sqrt{\frac{m}{2 \pi}} \frac{1}{\sqrt{2}}  \left[\lambda_{A }\left(x^\mu\right) e^{i m y}+c \cdot c \cdot\right]
\end{align}
where we find $\lambda_{\mu} \ \text{and} -i \lambda_{y} := \lambda $\  as the gauge parameters in $D$ dimensions. \\

The gauge symmetry of the Lagrangian in eq.(\ref{2.1}) in $D+1$ dimensions, becomes St{\"u}ckelberg symmetry of the KK reduced Lagrangian in D dimension:-
\begin{align}
    \delta h_{\mu \nu } &=  2!\partial_({_\mu} \lambda_{\nu)} - 2m \lambda \eta_{\mu \nu} , \\
    \delta B_{\mu} &= \partial{_\mu} \lambda + m \lambda_{\mu}, \label{2.10} \\
    \delta \phi &= 2m \lambda \label{2.11}
\end{align}
After performing the KK dimensional reduction to our gauge fixed Lagrangian in eq.(\ref{2.5}) in  $D+1$ dimensions, our Lagrangian becomes:-
 \begin{equation}
     \begin{split}
    c^{-1} \mathcal{L} & =\frac{1}{2} h_{\mu \nu }\left(\square-m^2\right) h^{\mu \nu} +  B_\mu\left(\square - m^2\right) B^\mu \\
    & + \frac{1}{2}\phi\left(\square-m^2\right) \phi - \frac{1}{4} \left( h - \phi \right) \left( \square - m^{2} \right) \left( h - \phi \right)
     \end{split}
 \end{equation}

With the following field redefinitions: 
\begin{align}
& h_{\mu \nu} \rightarrow h_{\mu \nu}+\left(\frac{1}{D-2}\right) \eta_{\mu \nu} \phi
\end{align}
We can cancel the mixed terms and find this diagonal Lagrangian:-
\begin{equation}
    \begin{split}
            & c^{-1} \mathcal{L}  =\frac{1}{2} h_{\mu \nu }\left(\square-m^2\right) h^{\mu \nu} 
 - \frac{1}{4} h \left(\square-m^2\right) h + \\
 & B_\mu\left(\square - m^2\right) B^\mu + \left(\frac{1}{2}- \frac{1}{(D-2)^{2}} + \frac{D}{2(D-2)^{2}} \right) \\
 & \phi\left(\square-m^2\right) \phi
    \end{split}
\end{equation}
For D=3, the Lagrangian becomes:-
\begin{equation}
    \begin{split}
            c^{-1} \mathcal{L} & =\frac{1}{2} h_{\mu \nu }\left(\square-m^2\right) h^{\mu \nu} 
 - \frac{1}{4} h \left(\square-m^2\right) h \\
& +  B_\mu\left(\square - m^2\right) B^\mu
 +  \phi\left(\square-m^2\right) \phi
    \end{split}
\end{equation}
 In 3D, the gauge fixing conditions are:-
 \begin{align}
     & \partial_{\mu} h^{\mu \nu} - \frac{1}{2} \partial^{\nu}h - m B^{\nu} =0 \label{2.20} \\
     & \partial_{\mu} B^{\mu} + \frac{m}{2} (h-2 \phi) = 0 \label{2.21}
 \end{align}
 By using the equations eq.(\ref{2.10}) and eq.(\ref{2.11}), we can set $ B_{\mu} $ and $\phi$ to zero. Therefore, from eq.(\ref{2.20}) and eq.(\ref{2.21}) we have:-
 \begin{align}
     \partial_{\mu} h^{\mu \nu} = 0 , \hspace{5mm} h = 0
 \end{align}
 Thus, we are left with a gauge fixed KK reduced Lagrangian for the two independent dof, which is of the following form:-
 \begin{align}
     c^{-1} \mathcal{L} = \frac{1}{2} h_{11}\left(\square-\left( mc \right)^2\right) h^{11} + \frac{1}{2} h_{12}\left(\square-\left( mc \right)^2\right) h^{12}
 \end{align}

Now, we can follow the same procedure as done in the Lagrangian density with projectors approach and Bergshoeff \cite{bergshoeff-2} - we make a complex combination of $h_{11}$ and $h_{12}$, use the expression in eq.(\ref{s2-100}) and take $c \longrightarrow \infty$ to find the Schr{\"o}dinger equation. \\

\textbf{Spin-3:} The Fronsdal Lagrangian for a massless real totally symmetric spin$=3$ field $\Phi_{ABC}$ in $D+1$ dimensions reads \cite{rakib} \cite{porrati}:

\begin{equation}
    \begin{split}
        & {c^{-1} \mathcal{L}}=-\frac{1}{2}\left(\partial_E \Phi_{A B C}\right)^2+\frac{3}{2}\left(\partial_A \Phi^{A B C}\right)^2+\frac{3}{2}\left(\partial_A \Phi_B^{\prime}\right)^2+ \\
        & 3 \ \Phi^{\prime C} \partial_A \partial_B \ \Phi^{A B}_{C} +\frac{3}{4}\left(\partial_A \ \Phi^{\prime A}\right)^2 
    \end{split}
\end{equation}

where $\Phi^{\prime}_{C} := \eta^{AB} \Phi_{ABC}$ is the trace of the totally symmetric spin$=3$ real field $\Phi_{ABC}$ . \\

The gauge symmetry of this Lagrangian is:-
\begin{align}
    \delta \Phi_{ABC} = 3! \ \partial_( {}_A \lambda_{BC)}  , \ \lambda^{A}_A = 0
\end{align}

The gauge fixing term for the D+1 dimensional Lagrangian reads:
\begin{align}
    c^{-1} \mathcal{L} &= - \left( \partial^{A}  \Phi_{A B C} - \frac{1}{2} \partial_{B} \Phi^{\prime}_{C} \right)^{2}
\end{align}

The KK dimensional reduction ansatz is:
\begin{align}
    \Phi_{A B C}\left(x^\mu, y\right)=  \sqrt{\frac{m}{2 \pi}} \frac{1}{\sqrt{2}}  \left[\varphi_{A B C}\left(x^\mu\right) e^{i m y}+c \cdot c \cdot\right]
\end{align}
where we compactify the $y :=x_{D+1}$ spatial dimension on a circle of radius = $1/m$. \\

We define the KK reduced fields in D dimensions as: $\varphi_{\mu \nu \rho} =: h_{{\mu \nu \rho}} , \ \varphi_{\mu \nu y} =: iW_{\mu \nu}, \
    \varphi_{\mu y y} =: -B_{\mu}, \
    \varphi_{y y y} =: -i \phi $. \\
    where, $h_{{\mu \nu \rho}}$ is a spin-$3$ field, $W_{\mu \nu}$ is a spin-$2$ field, $B_{\mu}$ is a spin-$1$ field and $\phi$ is a spin-$0$ field in dimension D.

We also write the KK ansatz for the gauge parameter $\Lambda_{AB}$ as:- 
\begin{align}
    \Lambda_{AB}\left(x^\mu, y\right)=  \sqrt{\frac{m}{2 \pi}} \frac{1}{\sqrt{2}}  \left[\lambda_{A B}\left(x^\mu\right) e^{i m y}+c \cdot c \cdot\right]
\end{align}
where we find $\lambda_{\mu \nu}, \lambda_{\mu} := i \lambda_{\mu y} \ \text{,and} \ \lambda :=\lambda_{yy}$ as the gauge parameters in $D$ dimensions. \\

The gauge symmetry of $D+1$ dimensions becomes St{\"u}ckelberg symmetry in D dimension:-
\begin{align}
    \delta h_{\mu \nu \rho } &= \partial_({_\mu} \lambda_{\nu \rho)} , \\
    \delta W_{\mu \nu} &= \partial_({_\mu} \lambda_{\nu )} + m \lambda_{\mu \nu}, \label{3.10} \\
    \delta B_{\mu} &= \partial_{\mu} \lambda + 2m \lambda_{\mu}, \label{3.11} \\
    \delta \phi &= 3m \lambda \label{3.12}
\end{align}

After performing the KK dimensional reduction, our Lagrangian becomes:-
\begin{equation}
    \begin{split}
        & c^{-1} \mathcal{L} = \frac{1}{2} h_{\mu \nu \rho} \square h^{\mu \nu \rho}+\frac{3}{2}\left(\partial_\mu h^{\mu \nu \rho}\right)^2+\frac{3}{4}\left(\partial_\mu h^\mu\right)^2 -\frac{3}{2} h_\mu \square h^\mu \\
        & +3 h_\rho \partial_\mu \partial_\nu h^{\mu \nu \rho}-\frac{m^2}{2} h_{\mu \nu \rho}^2+ \frac{3}{2} m^2 h_{\mu}^2-3 B_{\rho}  \partial_\mu \partial_\nu h^{\mu \nu \rho} \\
        & +3 B_{\rho} \square h^{\rho} + \frac{3}{2} B_{\rho} \left(\partial^\rho \partial^\mu h_\mu \right)+\frac{9}{4}(\partial \cdot B)^2 + \frac{3}{2} W_{\mu \nu} \square W^{\mu \nu} \\
        &+ 3 (\partial_{\mu} W^{\mu \nu})^2 - \frac{3}{2} W \square W + 3 W \partial_{\mu} \partial_{\nu} W^{\mu \nu} \\
        & -3 \phi (\partial_{\mu} \partial_{\nu} W^{\mu \nu}) + 3 \phi \square W - \phi \square \phi + 3 m h_{\mu \nu \rho} (\partial^{\mu} W^{\nu \rho}) -\\
        & 6 m h_{\rho} \partial_{\nu} W^{\nu \rho} - \frac{3}{2} m (\partial^{\mu} h_{\mu}) W +\frac{3}{2} m\left(\partial^\mu h_\mu\right) \phi- \\
        & \frac{9}{2} m\left(\partial^\mu B_\mu\right) W + \frac{3}{2} m\left(\partial^\mu B_\mu\right) \phi \\
        & + \frac{9 m^2}{4} W^2-\frac{3 m^2}{2} W \phi+\frac{m^2}{4} \phi^2 
        \label{h}
    \end{split}
\end{equation}

The gauge-fixing terms in D dimensions read:-
\begin{equation}
    \begin{split}
        & c^{-1} L_{gf1}=-\frac{3}{2} (\partial_\rho h^{\rho \mu \nu}-\frac{1}{2}\left(\partial^\mu h^\nu+\partial^\nu h^\mu\right) \\
        & -m W^{\mu v}+\left(\frac{m}{3}\right) \eta^{\mu v} W )^2 \label{3.13}
    \end{split}
\end{equation}
\begin{align}
       & c^{-1} L_{gf2}=-3\left(\partial_\nu W^{\mu v} -\frac{1}{2} \partial^\mu W-\left(\frac{m}{2}\right) h^\mu-\left(\frac{4 m}{3}\right) B^\mu\right)^2 \label{3.14} \\
       & c^{-1} L_{gf3} =-2 \left(\partial \cdot B-m W-3 m \phi \right)^2 
\end{align}

With the gauge fixing terms and the following field redefinitions:-
\begin{align}
    & h_{\mu \nu \rho} \rightarrow h_{\mu \nu \rho}+\frac{1}{D}\left(\eta_{\mu \nu} B_\rho+\eta_{\mu \rho} B_\nu+\eta_{\nu \rho} B_\mu\right) \\
& W_{\mu \nu} \rightarrow W_{\mu \nu}+\left(\frac{1}{D-2}\right) \eta_{\mu \nu} \phi
\end{align}
we can eliminate all the mixing terms in eq.(\ref{h}). Henceforth, we arrive at a Lagrangian, in which all kinetic terms are diagonal in D$=3$:-
\begin{equation}
    \begin{split}
         c^{-1} \mathcal{L} & =\frac{1}{2} h_{\mu \nu \rho}\left(\square-m^2\right) h^{\mu \nu \rho}-\frac{3}{2} h_\mu\left(\square - m^2\right) h^\mu \\
         & +\frac{3}{2} W_{\mu \nu}\left(\square-m^2\right) W^{\mu \nu} +2 B_\mu\left(\square-m^2\right) B^\mu \\
&-\frac{3}{4} W\left(\square-m^2\right) W+2 \phi\left(\square-m^2\right) \phi
\label{3.2}
    \end{split}
\end{equation}
 By using the gauge freedom given by eq.(\ref{3.10}), eq.(\ref{3.11}), and eq.(\ref{3.12}), we can set $W_{\mu \nu}$, $B_{\mu}$ and $\phi$ to zero. Henceforth, the gauge fixing conditions stemming from eq.(\ref{3.13}) and eq.(\ref{3.14}) implies:-
\begin{align}
    &\partial^{\mu} h_{\mu \nu \rho} = 0 ,\hspace{3mm} h^{\mu} = 0
\end{align}
So, we are left with a gauge fixed KK reduced Lagrangian for the two independent dof's of the spin-$3$ field in 3D:-
\begin{align}
     c^{-1} \mathcal{L} & =\frac{1}{2} h_{111}\left(\square- \left( mc \right)^2\right) h^{111} + \frac{1}{2} h_{112}\left(\square-\left( mc \right)^2\right) h^{112}
\end{align}
Now, we follow a similar procedure as described for spin-$2$ and find the Schr{\"o}dinger equation. \\

\textbf{Spin-$3/2$}: The Fronsdal Lagrangian for a massless complex  spin$=3/2$ field $\Psi_{A}$ in $D+1$ where, $D$ is even, dimensions reads:
\begin{align}
    c^{-1} \mathcal{L} = -i \ \Bar{\Psi}_{A} \gamma^{ABC} \partial_{B} \Psi_{C}
\end{align}

The gauge symmetry of this Lagrangian is:-
\begin{align}
    \delta \Psi_{A} = \partial_{A} E
\end{align}
 where, $E$ is a spinor which is the gauge parameter. \\
The KK dimensional reduction ansatz is:
\begin{align}
    \Psi_{A}\left(x^\mu, y\right)=  \sqrt{\frac{m}{2 \pi}} \psi_{A}\left(x^\mu\right) e^{i m y} 
\end{align}
where we compactify the $y :=x_{D+1}$ spatial dimension on a circle of radius = $1/m$. \\

We define the fields in D dimensions as: $\Psi_{\mu} =: -i \psi_{{\mu}} ,
    -\Psi_{y} =: \chi $.
    Here $\psi_{\mu}$ is a spin-$3/2$ field and $\chi$ is a spin-$1/2$ field in dimension D. \\
    
We also write the KK ansatz for the gauge parameter $E$ as:- 
\begin{align}
   E \left(x^\mu, y\right)=  \sqrt{\frac{m}{2 \pi}}  \varepsilon \left(x^\mu\right) e^{i m y} 
\end{align}
where we find $\varepsilon$\  as the gauge parameter in $D$ dimensions. \\

The gauge symmetry of the $D+1$ dimensional Lagrangian becomes St{\"u}ckelberg symmetry of the D dimensional Lagrangian:-
\begin{align}
    \delta \psi_{\mu} &= \partial_{\mu} \varepsilon \\
    \delta \chi &= m \varepsilon \label{3/2.1}
\end{align}
 After performing the KK dimensional reduction, our Lagrangian becomes:-
 \begin{equation}
     \begin{split}
        &c^{-1} \mathcal{L} = -i \Bar{\psi}_{\mu} \gamma^{\mu \nu \rho} \partial_{\nu} \psi_{\rho} - im \Bar{\psi}_{\mu} \gamma^{\mu \nu} \psi_{\nu} \\
        & + i \Bar{\psi}_{\mu} \gamma^{\mu \nu } \partial_{\nu} \chi + i \Bar{\chi} \overleftarrow{\partial}_{\mu} \gamma^{\mu \nu}  \psi_{\nu}
     \end{split}
 \end{equation}
 With the following field redefinition: 
\begin{align}
& \psi_{\mu} \rightarrow \psi_{\mu}+\left(\frac{1}{D-2}\right) \gamma_{\mu} \chi
\end{align}
we cancel the mixed kinetic terms and find:-
\begin{equation}
    \begin{split}
        & c^{-1} \mathcal{L} = -i \Bar{\psi}_{\mu} \gamma^{\mu \nu \rho} \partial_{\nu} \psi_{\rho} - \frac{(D-1)}{(D-2)} \Bar{\chi} \slashed{\partial} \chi - im \Bar{\psi}_{\mu} \gamma^{\mu \nu} \psi_{\nu} \\
        & - im \frac{(D-1)}{(D-2)} \Bar{\psi}_{\mu} \gamma^{\mu} \chi + im \frac{(D-1)}{(D-2)} \Bar{\chi} \gamma^{\mu} \psi_{\mu}  \\
        & + im \frac{D(D-1)}{(D-2)^{2}} \Bar{\chi} \chi
    \end{split}
\end{equation}
In D=3, the Lagrangian becomes:-
\begin{equation}
    \begin{split}
        c^{-1} \mathcal{L} &= -i \Bar{\psi}_{\mu} \gamma^{\mu \nu \rho} \partial_{\nu} \psi_{\rho} - 2 \Bar{\chi} \slashed{\partial} \chi - im \Bar{\psi}_{\mu} \gamma^{\mu \nu} \psi_{\nu} \\
        & - 2im \Bar{\psi}_{\mu} \gamma^{\mu} \chi + 2im  \Bar{\chi} \gamma^{\mu} \psi_{\mu}  + 6im \Bar{\chi} \chi
    \end{split}
\end{equation}
The equation of motion for $\chi$ is:-
\begin{align}
    -2 \slashed{\partial} \chi + 2im  \gamma^{\mu} \psi_{\mu}  + 6im \chi = 0
    \label{3/2.10}
\end{align}
Now, we can use the gauge freedom eq.(\ref{3/2.1}) to set $\chi^{\prime} $ to zero.
\begin{align}
     \chi^{\prime} &= \chi + \delta \chi = 0 \\
     \Rightarrow \varepsilon &= - \chi/ m 
\end{align}
We write eq.(\ref{3/2.10}) for the gauge transformed fields $\psi^{\prime}_{\mu}$ and $\chi^{\prime}$ and use $\chi^{\prime}=0$ to get:-
\begin{align}
     \gamma^{\mu} \psi^{\prime}_{\mu} &= 0 \\
     \gamma^{\mu} \psi_{\mu} + \gamma^{\mu} \partial_{\mu} \varepsilon &= 0 \\
\end{align}
Thus, we find the gauge condition:-
\begin{align}
    \gamma^{\mu} \psi_{\mu} = \slashed{\psi} = 0
\end{align}
by demanding:- 
\begin{align}
    \gamma^{\mu} \partial_{\mu} \varepsilon &= 0 \\
    \Rightarrow \square \varepsilon &= 0
\end{align}
This equation has non-trivial solutions. Therefore we still have some residual gauge freedom to transform our fields. We completely fix the gauge by setting:-
\begin{align}
    \psi_{0} = 0
\end{align}
Henceforth, we are left with a gauge fixed KK reduced Lagrangian for the independent dof $\psi_{1}$ of the spin-$3/2$ field in D$=3$:-
\begin{align}
    c^{-1} \mathcal{L} = -i \Bar{\psi}_{1} \slashed{\partial} \psi^{1} + im \Bar{\psi}_{1} \psi^{1}
\end{align}
From this Dirac Lagrangian, we obtain the Dirac equation, which imples that $\psi_{1}$ satisfies the KG equation. In the KG equation, we can take the $c \longrightarrow \infty$ limit and find the Schr{\"o}dinger equation. 

\section{Conclusion} 
We have shown how to take the non-relativistic limit and find the planar SE for higher spin bosonic and fermionic fields. We have demonstrated the procedure for both the equation of motion approach and action/Lagrangian density approach. We summarize the workings and results found in section-II, section-III and section-IV:-
\begin{itemize}
    \item In section-II, We worked at the level of equations of motion for both the bosonic and fermionic higher spin fields. We started from a massless theory in 4D. We used the light-cone gauge condition. From the equation of motion and light-cone gauge condition, we found that the field satisfies transversality and tracelessness conditions. By using the constraints, we were able to identify the two independent DoFs. They satisfy a wave equation. Then we defined a projective representation and used the null reduction ansatz of Bergshoeff et al. \cite{bergshoeff-2} in the wave equation. Thus we find the planar SE for the respective higher spin field.
    \item In section-III, we worked at the level of Lagrangian density for both the bosonic and fermionic higher spin fields. We used the projectors constructed by Kuzenko et al \cite{kuzenko} on the respective higher spin fields, to find the transverse, traceless field for the bosonic higher spin fields and transverse, gamma-traceless field for the fermionic higher spin fields. For the bosonic higher spin fields, we use the trasversality and tracelessness conditions to identify the independent DoFs and find the Lagrangian density for the independent DoFs. Then we define a complex field by combining the two DoFs. We take the projective representation of the complex field and substitute it in the Lagrangian density. Whereupon we take the non-relativistic limit $c \rightarrow \infty$ and find the non-relativistic Lagrangian density that produces, via the Euler-Lagrange equation, the planar SE for the respective bosonic higher spin field.
    For the fermionic higher spin fields, we used the transversality and gamma-tracelessness conditions to find the Dirac Lagrangian density for the respective transverse, gamma-traceless field. Then from the Dirac Lagrangian density via the Euler-Lagrange equation, we find that the transverse, gamma-traceless field satisfies the Dirac equation. Hence it also satisfies the KG equation. Then we take the projective representation of the transverse, gamma-traceless field and substitute it in the KG equation. After that, we take
    $c \rightarrow \infty$ and find the planar SE for the respective fermionic higher spin field.
    \item In section-$IV$, we used KK reduction to reduce a D+1 dimensional Lagrangian for a massless bosonic and fermionic field to a D dimensional Lagrangian. Then we gauge fixed the Lagrangian and took $c \longrightarrow \infty$ limit to find the planar Schr\"odinger equation. This method can also be systematically generalized to arbitrary higher spin bosonic and fermionic fields.
    \item This work establishes the existence of Schr\"odinger equation for arbitrary higher spin bosonic and fermionic fields in 3 dimensional Minkowski spacetime. This may be useful for condensed matter systems and other fields of research where researchers are interested in 3 dimensions and non-relativistic settings. There are a few possible extensions to this research. We may try to find SE for AdS spacetime\cite{hutchings} which may be useful for non-relativistic AdS3/CFT2. Another suggestion is mentioned to the author by Prof. P. K. Townsend that the non-relativistic limit of spin-$3/2$ and spin-$2$ fields may come from a supergravity theory in $3$D. We are currently working on these possible extensions. 
\end{itemize}

\vspace{1mm}

\acknowledgments
The author is thankful to Dr. Rakibur Rahman for suggesting the project and for providing important ideas for the project.
Part of the work is done in University of Dhaka and BRAC University. We are thankful to the respective institutions for their support.


\bibliographystyle{unsrt}
\bibliography{references}

\end{document}